\documentclass{iaus}
\usepackage{graphicx}

\def\MSUN{\rm M_{\odot}}
 
\def\MSUNYR{\rm M_{\odot}\,yr^{-1}}

\def\MDOT{\dot{M}}

\newbox\grsign \setbox\grsign=\hbox{$>$} \newdimen\grdimen \grdimen=\ht\grsign
\newbox\simlessbox \newbox\simgreatbox
\setbox\simgreatbox=\hbox{\raise.5ex\hbox{$>$}\llap
     {\lower.5ex\hbox{$\sim$}}}\ht1=\grdimen\dp1=0pt
\setbox\simlessbox=\hbox{\raise.5ex\hbox{$<$}\llap
     {\lower.5ex\hbox{$\sim$}}}\ht2=\grdimen\dp2=0pt

\title[Gas Accretion in  Quasars] 
{Dynamics of radiatively inefficient flows accreting onto 
radiatively efficient black hole objects}

\author[D. Proga]   
{Daniel Proga$^1$}

\affiliation{$^1$Department Physics, University of Nevada, Las Vegas,
Las Vegas, NV 89154, USA \break email: dproga@physics.unlv.edu\\
}

\pubyear{2006}
\volume{238}  
\pagerange{001--999}
\date{??? and in revised form ???}
\jname{Black Holes: from Stars to Galaxies -- across the Range of Masses}
\editors{V. Karas \& G. Matt, eds.}
\begin{document}

\maketitle

\begin{abstract}
I present results from numerical simulations of gas dynamics outside luminous 
accretion disks in active galactic nuclei. The gas, gravitationally captured 
by a super massive black hole, can be driven away by the energy and momentum
of the radiation emitted during black hole accretion.
Assuming axisymmetry,
I study how the mass accretion and outflow rates, and the flow dynamics 
respond to changes in radiation heating relative to radiation pressure.
I find that for a $10^8~\MSUN$ black hole with the accretion luminosity
of 0.6 of the Eddington luminosity the flow settles 
into a steady state and has two components:
(1) an equatorial inflow and
(2) a bipolar inflow/outflow with the outflow leaving the system
along the disk rotational axis. The inflow is a realization
of a Bondi-like accretion flow. The second component
is an example of a non-radial accretion flow which 
becomes an outflow once it is pushed close to the rotational axis 
where thermal expansion and radiation pressure accelerate it outward.
The main result of this preliminary work is 
that although the above two-component solution 
is robust, its properties are sensitive to
the geometry and spectral energy distribution of the radiation field.

\keywords{accretion, accretion disks, methods: numerical, hydrodynamics}
\end{abstract}

\firstsection 
\section{Introduction}

The radiation properties of active galactic nuclei (AGN)  and 
the AGN central location in their host galaxies imply that they play 
a very important role in determining the ionization structure and dynamics
of matter not only in their vicinity but also on larger,
galactic and even intergalactic scales
(Cotti \& Ostriker, 1997, 2001; King 2003; Murray, Quataert, \& Thompson 2005;
Sazonov et al. 2005; Springel, Di Matteo \& Hernquist 2005;
Hopkins et al. 2005, and references therin). Many observational
results support this suggestion, in particular
the presence of broad emission and absorption lines in AGN spectra.
The ionization structure and dynamics of the gas responsible for these lines
can be driven by radiation, even for sub-Eddington sources. 
The driving can be due to radiation pressure or radiation heating, 
or both (e.g., 
Begelman, McKee and Shields, 1982;
Shlosman, Vitello \& Shaviv 1985; 
Ostriker, McKee, \& Klein 1991;
Arav \& Li 1994;
Murray et al. 1995; 
Proga, Stone \& Kallman 2000;  
Proga \& Kallman 2002, 2004). 

In this paper, I present results from hydrodynamical simulations
of a non-rotating gas on sub-parsec- and parsec-scales in AGNs. I use
a simplified version of
the numerical method developed by Proga, Stone \& Kallman (2000)
to study a related problem of radiation driven disk winds in AGN
(for details see Proga et al. 2006 in preparation).
I consider an axisymmetric flow accreting onto a supermassive black hole (BH). 
The flow is non-spherical because it is irradiated 
by an accretion disk.
The disk radiation flux is the highest along 
the disk rotational axis and is gradually decreasing 
with increasing polar angle, $\theta$ as $\cos{\theta}$.
The flow is also irradiated by an isotropic corona (see eq. 2.1).
I take into account the radiation heating and cooling, radiation pressure due
to the electron scattering and spectral lines.
I adopt a simplified treatment of photoionization,
and radiative cooling and heating allowing for
a self-consistent calculation of 
the ionization state, and therefore the line force, 
in the flow.

\section{Results}

I assume the mass of the non-rotating BH, $M_{\rm BH}~=~10^8~\rm \MSUN$ and
the disk inner radius, $ r_\ast=~3~r_{\rm S}~=~8.8~\times~10^{13}$~cm 
throughout this paper. I consider the case with 
the rest mass conversion efficient 
$\eta=~0.0833$ and the mass accretion rate,
$\MDOT_{\rm a}$ =~10$^{26}~{\rm g~s^{-1}}$~(= 1.6~$\MSUNYR$).
These system parameters yield the accretion luminosity, 
$L = 7.5\times10^{45}~{\rm erg~s^{-1}}$, corresponding
to 0.6 of the Eddington luminosity.
To determine the radiation field, I specify
the fraction  of $L$ in the UV and X-ray band,
as $f_{\rm UV}$ and $f_{\rm X}$, respectively. 

Although, a quasar is powered by accretion, the disk accretion rate, 
that determines radiation, is not coupled to the rate, 
$\MDOT_{\rm in}(r_{\rm i})$  at which the non-rotating gas leaves 
the computational domain of these simulations
through the inner boundary.
This $\MDOT_{\rm a}$--$\MDOT_{\rm in}(r_{\rm i})$ decoupling 
is physically motivated because an accretion disk
is build from rotating gas but non-rotating gas will not significantly 
contribute to the disk mass and to the system luminosity.
In this sense, I study radiatively inefficient flows
accreting onto an object with a radiatively efficient accretion disk.

I present here preliminary results from simulations where all model parameters
are fixed except for $f_{\rm UV}$  and $f_{\rm X}$.
I consider three cases: case A with $f_{\rm UV}=0.5$  and $f_{\rm X}=0.5$, 
case B with $f_{\rm UV}=0.8$  and $f_{\rm X}=0.2$, and
case C with $f_{\rm UV}=0.95$  and $f_{\rm X}=0.05$
(see Table 1 for summary of the runs). 

At large radii, the radial radiation
force from the disk and spherical corona can be approximated as
\begin{equation}
{F}_r^{\rm rad}~(r,\theta)=
\frac{\sigma_e L}{4\pi r^2 c}\left[
2\cos{\theta}f_{\rm UV} (1+M(t))+f_{\rm X}\right],
\end{equation}
where $r$ and $\theta$ are the radius and polar angle in the spherical
polar coordinate system, respectively while the terms in the brackets 
with $f_{\rm UV}$ and $f_{\rm X}$
correspond to the disk and corona contribution, respectively. 
For simplicity, I assume that all UV photons are emitted by the disk
whereas all X-rays are emitted by
the corona.
The $M(t)$ is the so-called force multiplier - the numerical factor which
parameterizes by how much spectral lines increase the scattering
coefficient compared to the electron scattering coefficient (Castor, Abbott \& Klein 1975).

\begin{figure}
\begin{picture}(280,370)
\put(130,190){\includegraphics{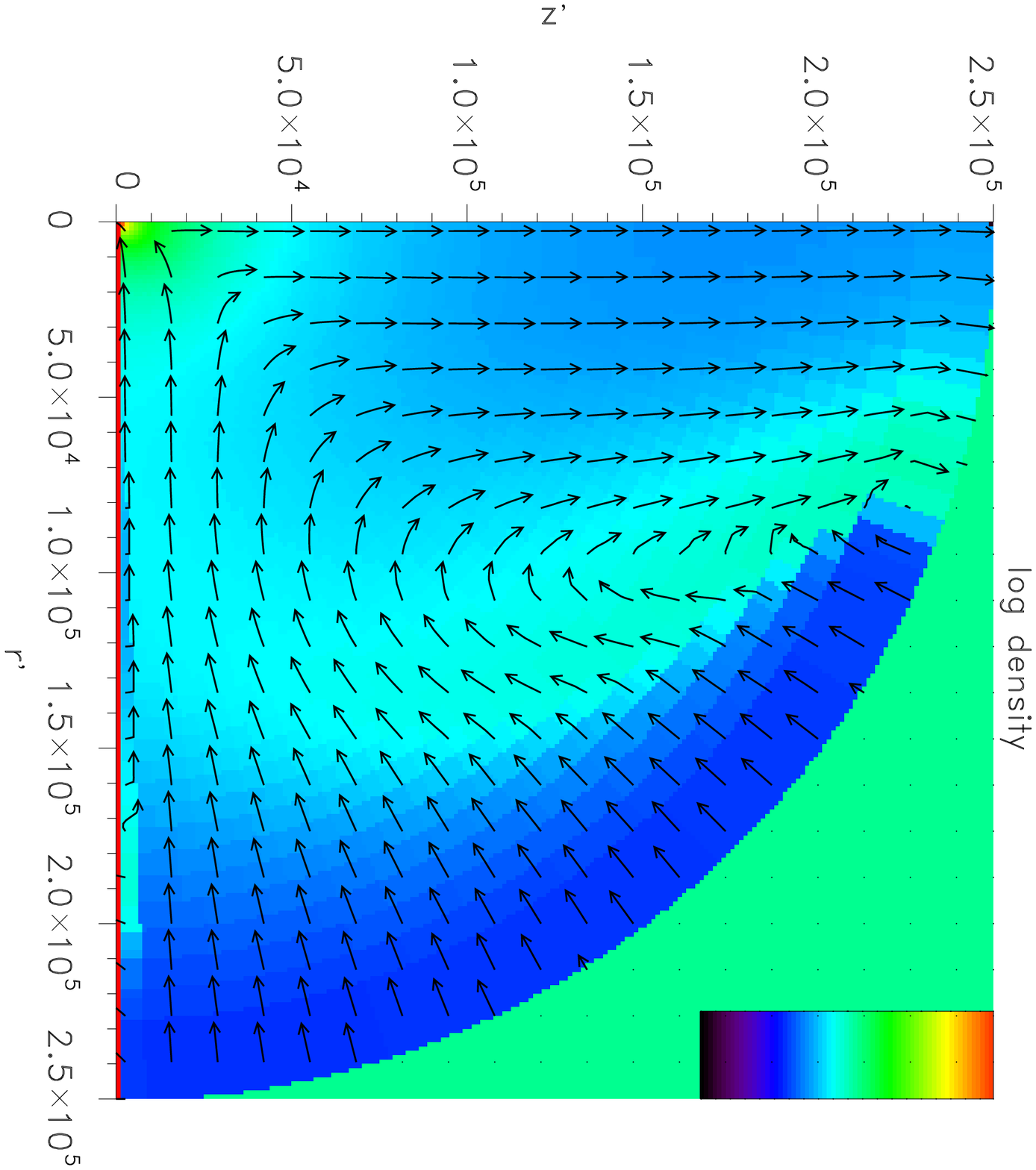}}
\put(310,190){\includegraphics{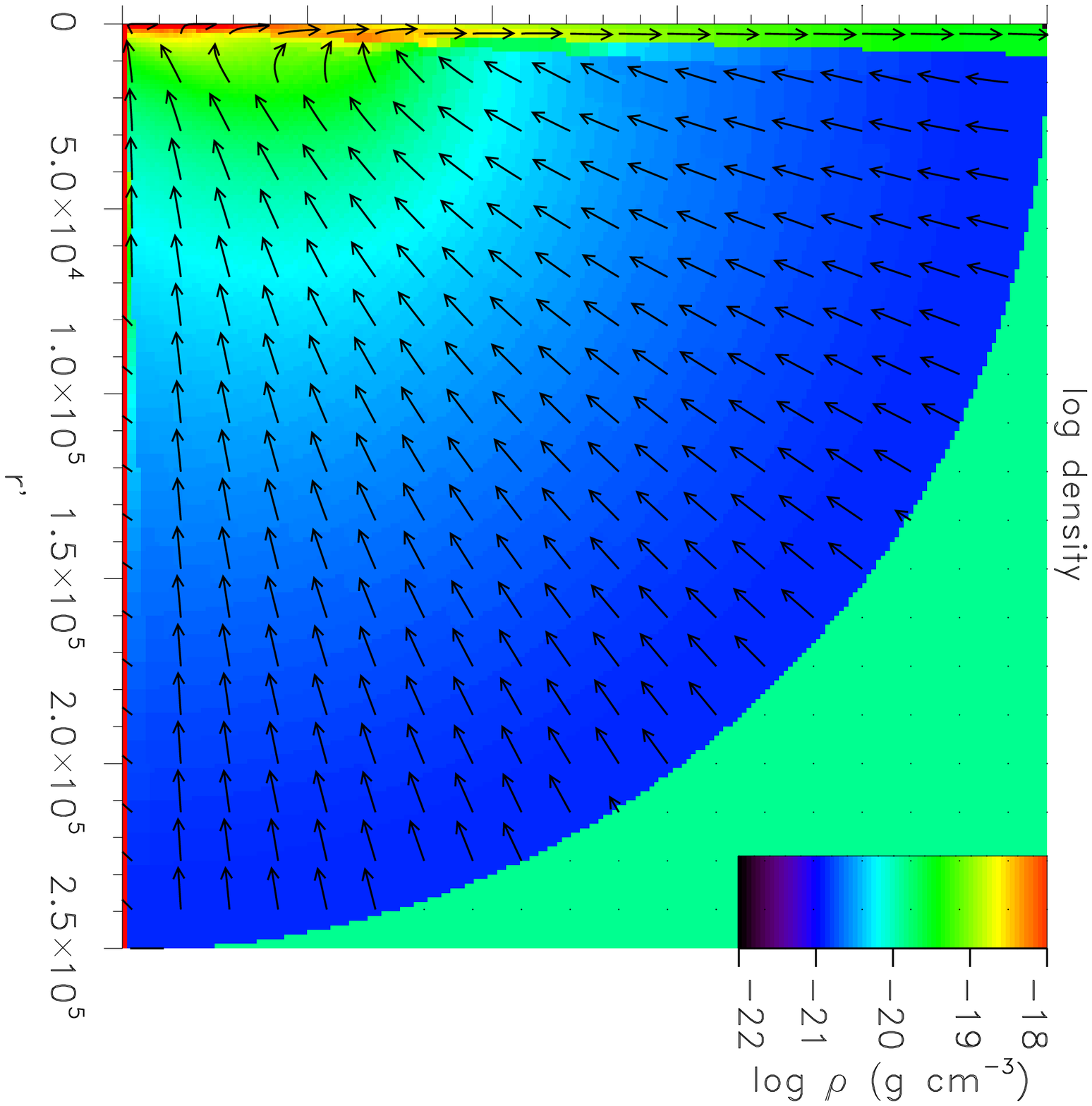}}

\put(130,10){\includegraphics{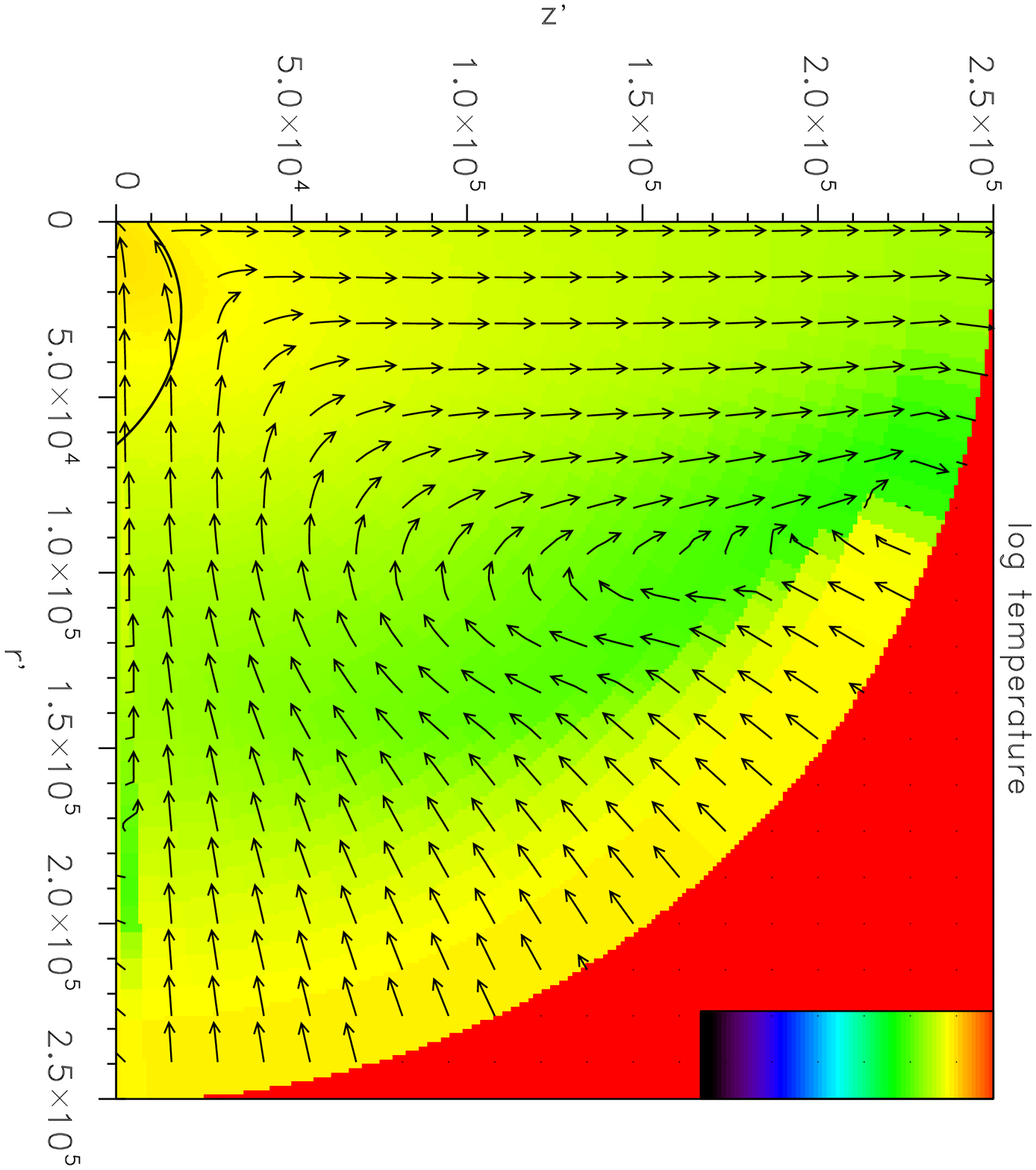}}
\put(310,10){\includegraphics{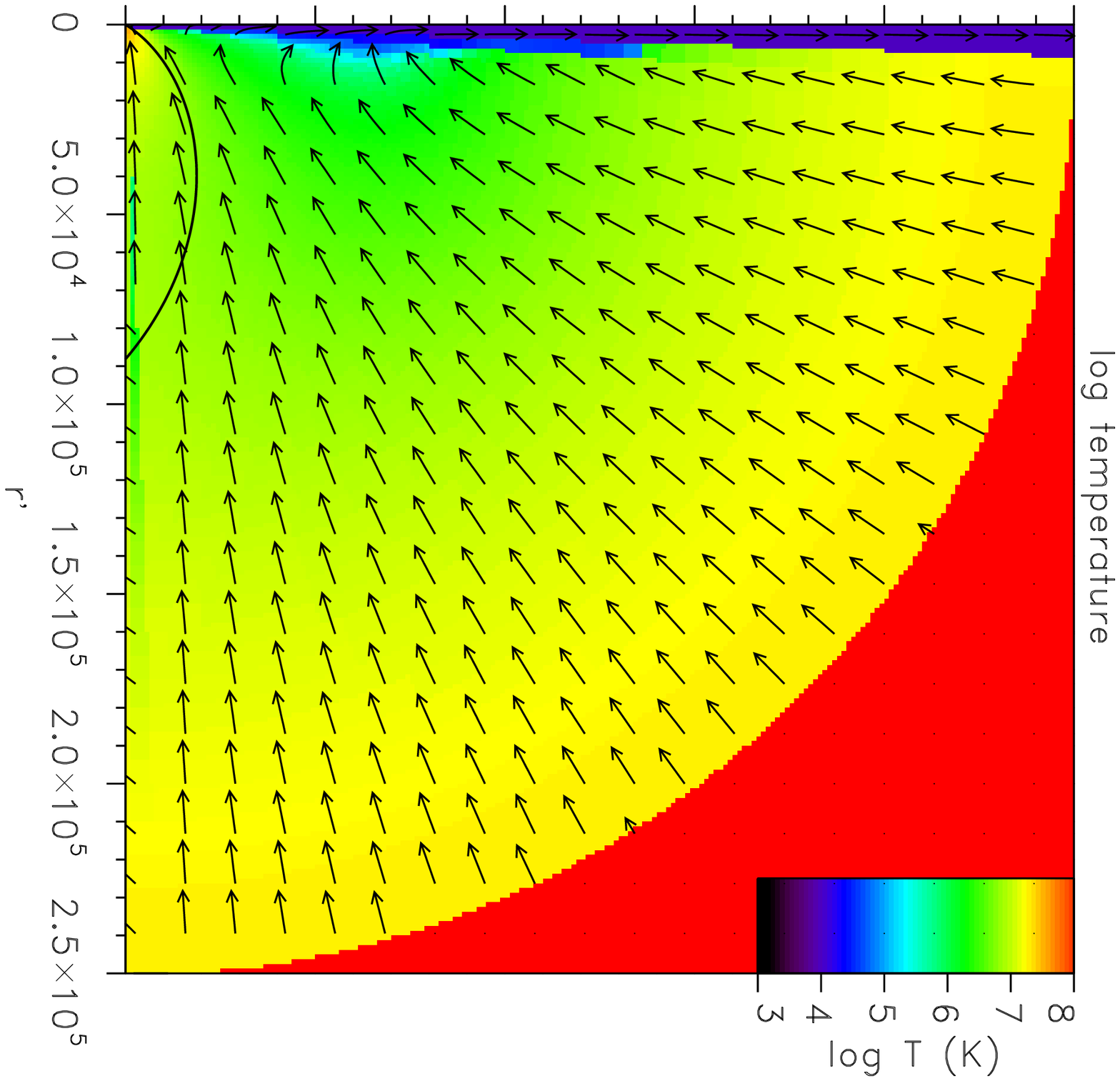}}
\end{picture}
\caption{ Comparison of the results for run A and C (left and
and right column, respectively).
{\it Top row of panels:}
Maps of logarithmic density overplotted by the direction of the 
poloidal velocity.
{\it Bottom row of panels:}
Maps of logarithmic temperature overplotted by the direction of the 
poloidal velocity. The solid curve in the bottom left corner
marks the position of the Compton radius corrected for the effects
of radiation pressure due to electron scattering (see eq. 2.2 in 
the main text).
The length scale is in units of the inner disk radius 
(i.e., $r' = r/r_\ast$ and $z' = z/r_\ast$).
The computational domain is defined to occupy
the angular range $0^o \leq \theta \leq 90^o$ and the radial range
$r_{\rm i}~=~500~r_\ast \leq r \leq \ r_{\rm o}~=~ 2.5~\times~10^5~r_\ast$.
}
\end{figure}

Figure 1 compares the results from run A and run C.
I specified the outer boundary in the following way.
The density and temperature at the outer radius, $r_{\rm o}$ was set to
$\rho_{\rm 0}=10^{-21}~\rm g~cm^{-3}$ and  
$T_{\rm 0}=2\times10^7$~K, respectively.
The velocity was set to zero.
At the outer radial boundary,
during the evolution of each model I continue to apply these constraints that
the density and temperature are fixed at constant values at all times.
For the initial conditions,
I set all variables constant and equal to their values at the outer boundary, 
as listed above. The figure shows the instantaneous density, temperature, and
distributions and the poloidal velocity field of the models.
Additionally, it also shows the so-called Compton radius
corrected for the effects of radiation pressure due to electron scattering
\begin{equation}
{\bar R}_{\rm C} \equiv R_{\rm C} [1-\Gamma (2 \cos{\theta} f_{\rm UV}+f_{\rm X})],
\end{equation}
where 
$R_{\rm C}\equiv G M_{\rm BH} \mu m_p/k T_{\rm C}=8.03\times10^{18}$~cm~
=$9.1\times10^4~r_\ast$ is the uncorrected Compton radius
for the Compton temperature, $T_{\rm C}=2\times 10^7$~K.

In all runs, the flow settles quickly into a steady state
(within $\sim$ a few $\times 10^{12}$~s
which correspond to a few dynamical time scales
at $r_{\rm o}$, $\tau=(r_o^3/G M_{\rm BH})^{1/2}=9 \times 10^{11}$~s ).
The steady state consists of two flow components
(1) an equatorial inflow and
(2) a bipolar inflow/outflow with the outflow leaving the system
along the pole. The outflow is collimated by the inflow.
The calculations capture the subsonic and supersonic
parts of both the inflow and outflow.
Although the same components can be identified in all runs,
their size, density and temperature, and the degree  of outflow
collimation depend on $f_{\rm UV}$ and $f_{\rm X}$ 
(the spectral energy distribution  and geometry of the radiation field).
In particular, the outflow power (e.g., measured
as the kinetic energy, $P_{\rm k}$, and 
thermal energy  $P_{\rm th}$ carried by the outflow) and the degree of
collimation is higher for the model with
the radiation dominated by the UV/disk emission (run C)
than for the model with the radiation dominated by the X-ray/corona emission
(run A).
I note that a very narrow outflow driven by radiation 
pressure on lines can carry more energy and mass than a broad outflow driven 
by thermal expansion (compared results for runs C and A).

\section{Conclusions}

The simulations show that AGN can have a substantial outflow which
originates from the inflow at large radii. Such an outflow
can control the rate at which non-rotating matter is supplied
to the AGN central engine because the outflow mass loss rate, 
$\MDOT_{\rm out}(r_{\rm i})$,
can be significantly higher than the mass inflow rate at small radii,
$\MDOT_{\rm in}(r_{\rm i})$.
For example, in run C,
as little as 10\% of the inflow at large radii  reaches small 
radii because 90\% of the inflow is turned into an outflow.
However, even the power of the strongest outflow 
is very low compared to the radiation power (i.e., for run C,
$P_{\rm k}/L=4\times 10^{-4}$). Finally, the inflows and outflows,
found in these simulations, can be related to material
responsible for broad absorption and emission lines,
and narrow absorption and emission lines observed in X-ray and UV 
spectra of AGN.

\begin{table}\def~{\hphantom{0}}
  \begin{center}
\caption{ Summary of results}
\begin{tabular}{l c c  c c  c c c c   } \\ \hline
        & & &     &            &                       &   &    \\
Run &  $f_{\rm UV}$ & $f_{\rm X} $ & $\MDOT_{\rm in}(r_{\rm o})$ & $\MDOT_{\rm in}(r_{\rm i})$  & $\MDOT_{\rm out}(r_{\rm o})$   &     $v_r$         & $P_{\rm k}({\rm r}_o)$        &  $P_{\rm t}({\rm r}_o)$        \\ \hline
A   & 0.5& 0.5& 4 &  1            & 3  & 700           & 2 & 4        \\
B   & 0.8& 0.2& 8 & 3            & 5  & 4000          & 100 & 1        \\
C   & 0.95 & 0.05 &  9 & 1            & 8  & 6700              & 300 & 0.03    \\ \hline
  \end{tabular}

The quantities in the table are in the following units:
$\MDOT_{\rm in}(r_{\rm o})$, $\MDOT_{\rm in}(r_{\rm i})$, and 
$\MDOT_{\rm out}(r_{\rm o})$ are in units of $10^{25}$~g~s$^{-1}$, 
$v_r$ is in units of $\rm km~s^{-1}$, 
and  $P_{\rm k}(r_{\rm o})$ and  $P_{\rm t}(r_{\rm o})$ are in units 
of $10^{40}$~erg~s$^{-1}$.
 \end{center}
\end{table}

\begin{acknowledgments}
This work is supported by NASA through grants 
and HST-AR-10305 and HST-AR-10680 
from the Space Telescope Science Institute, 
which is operated by the Association of Universities for Research 
in Astronomy, Inc., under NASA contract NAS5-26555.
\end{acknowledgments}

\end{document}